# An Empirical Study of UDP (CBR) Packet Performance over AODV Single & Multi-Channel Parallel Transmission in MANET

Md. Monzur Morshed* [1, 2], Meftah Ur Rahman[2, 3], Md. Rafiqul Islam[1]
{m.monzur@gmail.com, monzur@tigerhats.org},
tasnimbd@gmail.com,
rafiqulislam@aiub.edu
Department of Computer Science
American International University-Bangladesh[1]
TigerHATS Research Team-Bangladesh[2]
George Mason University-USA[3]

*Abstract*- **Mobile Ad-hoc Network is a temporary network which is the cooperative engagement of a collection of standalone mobile nodes that are not connected to any external network. It is a decentralized network where mobile nodes can be easily deployed in almost any environment without sophisticated infrastructure support. An empirical study has been done for AODV routing protocol under single channel and multi channel environment using the tool NS2. To compare the performance of AODV in the two environments, the simulation results have been analyzed by graphical manner and trace file based on QoS metrics such as throughput, packet drop, delay and jitter. The simulation result analysis verifies the AODV routing protocol performances for single channel and multi channel. After the analysis of the simulation scenario we suggest that use of Parallel MAC (P-MAC) may enhance the performance for multi channel.**

## I. Introduction

Mobile Ad-hoc networks (MANETs) have the capability to establish networks at anytime, anywhere holding the promise of the next generation communication features. With no pre-existing communications infrastructure, MANET represents such complex distributed system that enables seamless networking. It is a composition of a group of mobile, wireless nodes which cooperate in forwarding packets in a multi-hop fashion without any centralized administration. In MANET, each mobile node acts as a router as well as an end node which is either source or destination. AODV is a widely used routing protocol for MANET offering quick adaptation to dynamic link conditions, low network utilization, low processing, memory overhead, and determines unicast routes to destinations within the ad hoc network. AODV is re-active routing protocol which determines a route to the destination only when packets are sent to destination. If the wireless nodes are within the range of one another, the routing is not necessary. If a node moves out of range then the node will not be able to communicate with others directly, intermediate nodes are needed to organize the network which takes care of the data transmission [1, 2].

## II. Ad-hoc On-demand distance vector (AODV)

Ad-hoc On-demand distance vector (AODV) is a variant of classical distance vector routing algorithm. Like DSDV, AODV provides loop free routes in case of link breakage but unlike DSDV, it doesn't require global periodic routing advertisement. AODV experiences unacceptably long waits frequently before transmitting urgent information because of its on demand fashion of route discovery [3]. It borrows the basic on-demand mechanism of Route Discovery and Route Maintenance from DSR protocol, plus the use of hop-by-hop routing, sequence numbers, and periodic beacons from DSDV protocol [4]. The AODV protocol is loop-free and avoids the count-to-infinity problem by the use of sequence numbers. Additionally, AODV forms trees which connect multicast group members. The trees are composed of the group members and the nodes needed to connect the members.

AODV protocol uses a simple request-reply mechanism for route discovery. When a node desires to send a packet to destination, it checks its routing table to determine if it has a current route to the destination. If yes, forwards the packet to next hop node. If No, it initiates a route discovery process.

| Source Address | Request ID | Destination Address | Source Sequence # | Dest. Sequence # | Hop count |
|---|---|---|---|---|---|

**Figure 1:** Format of a Route Request (RREQ) packet

From Figure 1, Route discovery process begins with the creation of a Route Request (RREQ) packet where source node creates it. Source node sends a Routes Request message to its neighbors. The packet contains source node's IP address, source node's current sequence number, destination IP address, destination sequence number. Packet also contains broadcast ID number. Broadcast ID gets incremented each time a source node uses RREQ. Broadcasting is done via Flooding. Source node's IP address and Request ID fields uniquely identify the ROUTE REQUEST packet to allow nodes to discard any duplicates they may receive. Sequence number of source and the most recent value of destination sequence number that the source has seen and the Hop count field will keep track of how many hops the packet has traveled. When source include destination sequence numbers in its route request that actually last known destination sequence number for a particular destination. Every

This research is supported by AIUB & TigerHATS Research Team.
For more information please visit www.aiub.edu, www.tigerhats.org

intermediate nodes store most recent sequence number of source. As a RREQ propagates through the network, intermediate nodes use it to update their routing tables. Once an intermediate node receives a RREQ, the node sets up a reverse route entry for the source node in its route table.

| Source Address | Destination Address | Destination Sequence # | Hop count | Lifetime |
|---|---|---|---|---|

**Figure 2:** Format of a Route Reply (RREP) packet

From Figure 2, using the reverse route a node can send a RREP (Route Reply packet) to the source. A node receiving the RREQ may send a route reply (RREP) if it is either the destination or if it has a route to the destination with corresponding sequence number greater than or equal to that contained in the RREQ. If a neighbor has a route to destination then it unicasts a RREP back to the source node. If neighbors have no route then it rebroadcast RREQ and increment hop count. When a RREQ reaches a destination node, the destination route is made available by unicasting a RREP back to the source route. The receiver looks up the destination in its route table.

If they receive a RREQ which they have already processed, they discard the RREQ and do not forward it. As the RREP propagates back to the source node, intermediate nodes update their routing tables. Source node set up forward pointers to the destination. Once the source node receives the RREP, it may begin to forward data packets to the destination. A node may receive multiple RREP for a given destination from more than one neighbor. The node only forwards the first RREP it receives. It may forward another RREP if that has greater destination sequence number or a smaller hop count, it may update its routing information for that destination and begin using the better route.

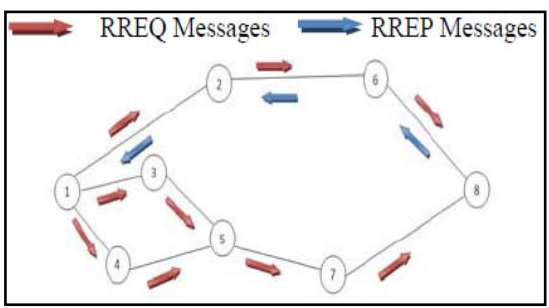

**Figure 3:** AODV mechanism

AODV uses sequence numbers to ensure the freshness of routes. To test freshness it compares destination sequence number, if RREQ packet destination sequence number is greater than the Route destination sequence numbers assumes route is still present and remains unused. As long as the route remains active, it will continue to be maintained. A route is considered active as long as there are data packets periodically traveling from the source to the destination along that path. Once the source stops sending data packets, the links will time out and eventually be deleted from the intermediate node routing tables. From figure: 2, if a link break occurs while the route is active, the node upstream of the break propagates a route error (RERR) message to the source node to inform it of the now unreachable destination(s). After receiving the RERR, if the source node still desires the route, it can reinitiate route discovery [5, 6].

## III. Single Channel vs. Multi-Channel

Among a great number of routing protocols, it is common to use a single-radio and single-path routing method, e.g. Dynamic Source Routing (DSR) [4], Ad-hoc on demand Distance Vector routing (AODV) [5]. In this case, packets travel along the chain of nodes toward their destinations and all nodes are working with one radio in the same channel, as shown in Figure 3. Successive packets on a single chain may interfere with each other causing channel competition and collision in the MAC layer. Ideally end-to-end throughput could achieve at most 1/3 of the effective MAC layer data rate, since at one time; among any three continuous nodes only one can make transmission in the same channel.

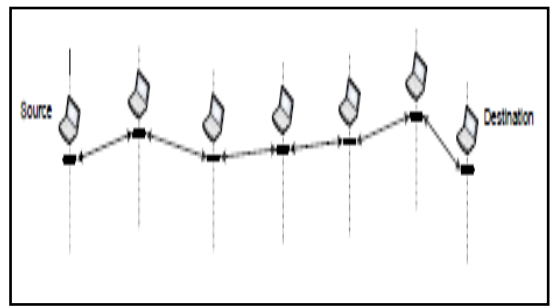

**Figure 4:** Single Channel Routing

Some researchers have proposed multi-channel/multi-radio solutions using more channels/radios to receive and send data in different channels simultaneously, such as [7], [8] and [9]. In this scenario, an ideal multi-channel/multi-radio routing protocol could help achieve end-to-end throughput almost as high as the effective MAC data rate.

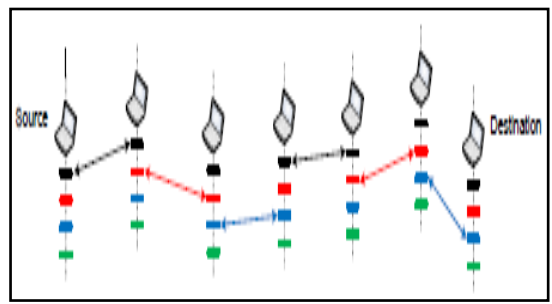

**Figure 5:** Multi channel routing

Considering the scenario in Figure.5, we assume that the MAC protocol can always select an appropriate radio and schedule perfectly. At time slot 1, the first node transmits the first packet to second node on channel 1. At time slot 2, they can transmit at the same time using different radios as well. If radio resources are enough for MAC protocol, every node can continuously inject one packet every time slot.

## IV. Parallel Transmission

Two interfaces working on different frequency can forward the packet simultaneously as it is regarded as no self-interference. To achieve more capacity, the utilization of diversity of multi-radio and multi-channel becomes a key point. Many channel assignment algorithms try to find the optimization assignment using both centralized and distributed methods. Multi-channel and multi-radio comes in order to decrease the self-interference under the transmission range and enhance throughput of end-to-end transmission [10].

## V. Simulation Topology

Our Simulation environment consists of 4 wireless mobile nodes which are placed uniformly and forming a Mobile Ad-hoc Network, moving about over a $1000 \times 1000$ meters area for 20 seconds of simulated time. We have used standard two-ray ground propagation model, the IEEE 802.11 MAC, and omni-directional antenna model of NS2. We have used AODV routing algorithm and interface queue length 50 at each node. The source nodes are respectively 0, 2 and the receiving nodes are respectively 1, 3. We have carried out the simulation in 4 separate cases such as Case 1, Case 2, Case 3 and Case 4.

**Case 1:** Single Channel, No Mobility
**Case 2:** Multi-Channel, No Mobility
**Case 3:** Single Channel with Mobility
**Case 4:** Multi-Channel with Mobility

For Case 1,2 the simulation run time is 20 sec long without node mobility where as mobile nodes 0,2 are source node and 1,3 are destination node.

For Case 3, 4 the simulation run time is 20 sec long and we have segmented the simulation time in two intervals. For first time interval of 0-15 sec there was no node mobility, only for last 5 sec the node mobility was enabled.

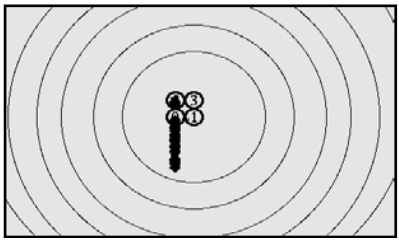

**Figure 6:** Simulation Topology for AODV single channel

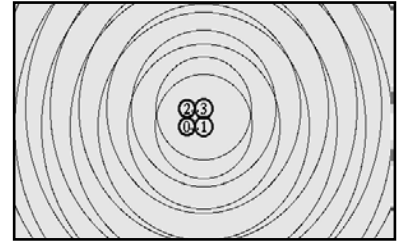

**Figure 7**: Simulation Topology for AODV multi channel

## VI. Simulation Description

**Table 1:** Simulation parameters

| Method | Value |
|---|---|
| Channel type | Channel/Wireless channel |
| Radio-propagation model | Propagation/Two ray round |
| Network interface type | Phy/wirelessphy |
| MAC type | Mac/802.11 |
| Interface queue type | Queue/Drop Tail |
| Link Layer Type | LL |
| Antenna | Antenna/omni antenna |
| Maximum packet in ifq | 50 |
| Area (m×m) | 1000×1000 |
| Number of mobile nodes | 4 |
| Source type | UDP (CBR) |
| Simulation Time | 20 sec |
| Routing protocol | AODV, AODV Multi-Channel |

## VII. QoS METRICS

We used different parameter of QoS metrics such as delay, jitter, packet drop and throughput to understand the behavior of AODV and AODV multi channel Routing Protocol.

## VIII. SIMULATION RESULT

*A. Drop*

The routers might fail to deliver (drop) some packets if they arrive when their buffers are already full. Some, none, or all of the packets might be dropped, depending on the state of the network, and it is impossible to determine what will happen in advance. The receiving application may ask for this information to be retransmitted, possibly causing severe delays in the overall transmission. Packet drop is equal to number of packets sent from source minus number of packet received in the path of destination e.g.;

*No of Packets Dropped = No of pkt Sent – No of pkt Received*

For Single & Multi Channel with mobility & no mobility AODV:

| Case | Total CBR sent | Total CBR received | Total CBR drop |
|------|----------------|--------------------|-----------------|
| Case 1 | 25002 | 12498 | 12408 |
| Case 2 | 25002 | 24635 | 367 |
| Case 3 | 25002 | 10298 | 14598 |
| Case 4 | 25002 | 22097 | 2879 |

*B. Throughput*

Throughput is the measurement of number of packets passing through the network in a unit of time. This metric show the total number of packets that have been successfully delivered to the destination nodes and throughput improves with increasing nodes density. Throughput can be defined by:

$$\frac{\Sigma\ Node\ Throughputs\ of\ Data\ Transmission}{Total\ number\ of\ nodes}$$

For Single channel:

| total sending throughput | total receiving throughput |
|--------------------------|----------------------------|
| 25002000 | 12742860 |

For multi channel:

| total sending throughput | total receiving throughput |
|--------------------------|----------------------------|
| 25002000 | 25127700 |

*C. Delay*

A specific packet is transmitting from source to destination and calculates the difference between send times and received times. Delays due to route discovery, queuing, propagation and transfer time are included in the delay metric.

*Packet Delay = packets receive time – packet send time*

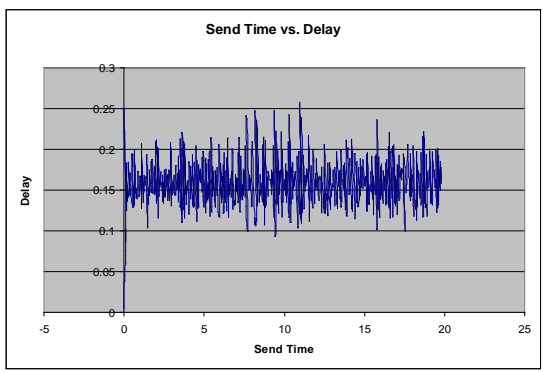

**Figure 8**: Delay graph for Single Channel AODV

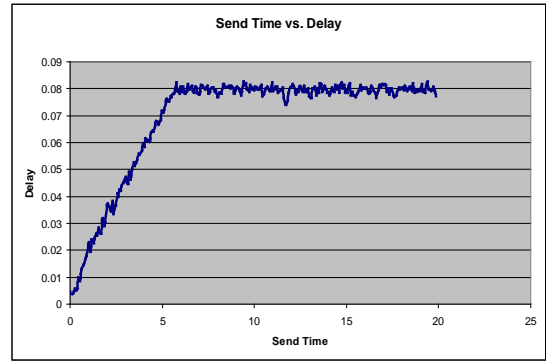

**Figure 9:** Delay graph for multi channel AODV

*D. Jitter*

Jitter is the variation in the time between packets arrival, caused by network congestion, timing drift, or route changes. In jitter calculation the variation in the packet arrival time is expected to minimum. The delays between the different packets need to be low if we want better performance in Mobile Ad-hoc Networks.

*Jitter ( i ) = Delay (i+1) – Delay (i)  where i =  1,2,3…..n*

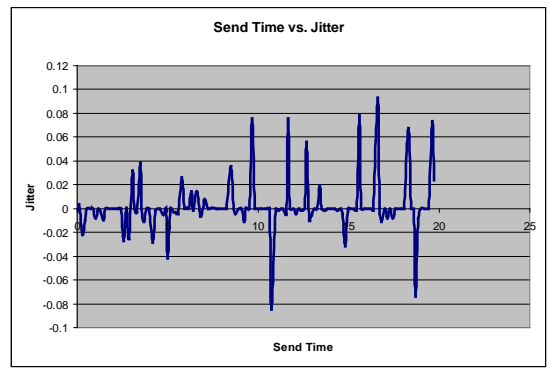

**Figure 10**: Jitter for Single channel AODV

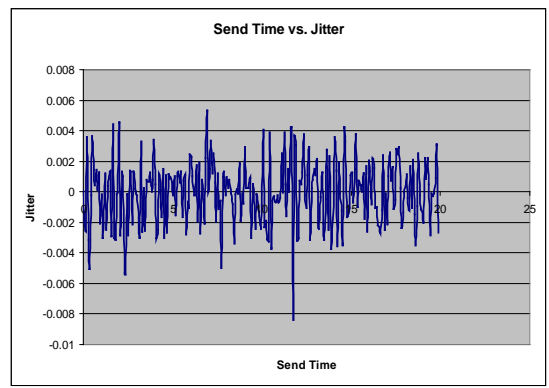

**Figure 11**: Jitter for multi channel AODV

## IX. RESULT ANALYSIS TABLE

|  | Case 1 | Case 2 | Case 3 | Case 4 |
|---|---|---|---|---|
| **Drop Rate** | 49.63% | 1.47% | 58.39% | 11.52% |
| **Packet Delivery Ratio** | 49.99% | 98.53% | 41.19% | 88.38% |

## X. CONCLUSION

We carried out our simulation work in 4 (four) cases and analyzed UDP (CBR) packets where as the packets drop rates for Case 1, Case 2, Case 2, Case 3, Case 4 are respectively 49.63%, 1.47%, 58.39%, 11.52%. For AODV routing protocol in single & multi channel, the packet delivery ratio is independent of offered traffic load, with both channels are respectively 49.99%, 98.53%**,** 41.19%, and 88.38% in all cases. So we can conclude that considering Cases 1,2,3,4 AODV indicating its highest efficiency and perform better for Case 2: Multi-Channel, No Mobility. Multi channel with no mobility gives better performance because we have used static nodes for Case 2 and the fact is there is no routing overhead for route maintenance and less congestion. While nodes communicate with other nodes in multi channel then intermediate nodes between source-destination should have similar configuration which means all the nodes between source-destination have to be homogenous. If the intermediate nodes between source-detonation use different communication channel then the communication will not take place. To avoid this problem, using of idea of P-MAC (Parallel MAC) may prove to be useful. Further research is necessary to properly harness the usability of P-MAC in Mobile Ad Hoc Networks [12].